\documentclass[11pt]{article}
\usepackage{epsf,amsmath,amssymb,graphicx,scalefnt}
\usepackage{rotating}
\newcommand{\abbrev}{\scalefont{.9}}
\newcommand{\ep}{\epsilon}
\newcommand{\bare}{{\rm B}}

\newcommand{\apib}{\frac{\alpha_s^\bare}{\pi}}

\newcommand{\mhiggs}{M_{\rm H}}
\newcommand{\mtop}{M_{\rm t}}
\newcommand{\eqn}[1]{Eq.\,(\ref{#1})}
\newcommand{\fig}[1]{Fig.\,\ref{#1}}

\newcommand{\dd}{{\rm d}}
\newcommand{\dotprod}[2]{#1\!\cdot\!#2}

\newcommand{\order}[1]{{\cal O}(#1)}
\newcommand{\lhc}{{\abbrev LHC}}
\newcommand{\qcd}{{\abbrev QCD}}

\newcommand{\lo}{{\abbrev LO}}
\newcommand{\nlo}{{\abbrev NLO}}
\newcommand{\nnlo}{{\abbrev NNLO}}

\date{}
\title{\vspace*{-6em}
  \begin{flushright}
    {\sf\small July 2009 --- WUB/09-08}
  \end{flushright}
  \vspace*{2em}
  Top mass effects in Higgs production at next-to-next-to-leading
  order QCD: virtual corrections}
\author{Robert V. Harlander and Kemal J. Ozeren\\[1em]
{\it Fachbereich C, Bergische Universit\"at Wuppertal}\\
{\it  42097 Wuppertal,  Germany}\\[.5em]
{\scalefont{.8}{\it email:} \parbox[t]{.4\textwidth}{
    {\tt robert.harlander@uni-wuppertal.de}\\ {\tt
    ozeren@physik.uni-wuppertal.de}}}}

\begin{document}
\maketitle
\begin{abstract}
Top quark mass suppressed terms are calculated for the virtual amplitude
for Higgs production in gluon fusion at three-loop level,
i.e.\ $\order{\alpha_s^3}$. The method of asymptotic expansions in its
automated form is used to evaluate the first three
non-vanishing orders in terms of $\mhiggs^2/\mtop^2$, where the first
order corresponds to the known results of the effective Lagrangian
approach.
\end{abstract}



\section{Introduction}

Radiative corrections to Higgs production through gluon fusion are known
to be unusually
large~\cite{Dawson:1990zj,Djouadi:1991tk,Graudenz:1992pv,Spira:1993bb}.
The inclusive next-to-next-to-leading order (\nnlo{}) cross section
$\sigma(pp/p\bar p\to H+X)$ exceeds the \lo{} prediction by roughly a
factor of two at \lhc{} energies, and even up to a factor of three at
the
Tevatron~\cite{Harlander:2002wh,Anastasiou:2002yz,Ravindran:2003um}. Recent
compilations of the currently available contributions to the production
cross section can be found in
Refs.\,\cite{deFlorian:2009hc,Anastasiou:2008tj}.

The current \nnlo{} prediction is based on the assumption that the top
mass dependence is largely determined by the \lo{} expression, while the
higher order terms can be evaluated in the limit of infinitely heavy top
mass
$\mtop$~\cite{Dawson:1993qf,Graudenz:1992pv,Spira:1993bb,Kramer:1996iq}. At
\nlo{}, where a comparison with the full mass dependence of the cross
section is possible, the heavy-top approximation is valid at the 2-3\%
level for Higgs masses $M_H<2M_t$ (see, e.g.,
Ref.\,\cite{Harlander:2003xy}). Even at $M_H\approx 1$\,TeV, the
deviation from the full \nlo{} result amounts to only about 10\%.

The fact that the heavy top limit works so well is at first
sight surprising, because it assumes that $\mtop$ is larger than any
other scale in the process. This is certainly not the case at the
\lhc{} with a prospected hadronic center-of-mass energy of
$\sqrt{s}=14$\,TeV. However, one can argue that since the cross section
is dominated by soft gluon radiation parton scatterings with
energies $\sqrt{\hat s}$ much larger than $2\mtop$ are strongly suppressed.

It is indeed observed that an expansion of the partonic cross section
$\hat\sigma$ in powers of $(1-z)$, where $z=\mhiggs^2/\hat s$,
converges rather quickly to the exact result~\cite{Harlander:2002wh}.
On the other hand, resummation of the soft terms does not lead to
a big effect at any of the three lowest orders in perturbation
theory~\cite{Catani:2003zt}. 

Recently it has been suggested that the size of the radiative
corrections is due to the transition from space- to time-like momenta,
and in fact, numerical studies show that the bulk of the radiative
corrections can be obtained by resumming the leading $\pi^2$-terms that
arise from this transition~\cite{Ahrens:2008qu:2008nc}.

These unresolved issues leave one with a certain amount of doubt as to
the use of the heavy-top limit at \nnlo{}. There is however surprisingly
little activity in the field that addresses the validity of this
approximation. Besides the \nlo{} calculations for the inclusive cross
section mentioned
before~\cite{Dawson:1993qf,Graudenz:1992pv,Spira:1993bb,Kramer:1996iq},
there are studies concerning the mass effects on differential
distributions~\cite{DelDuca:2001eu:2001fn:2003ba,
  Keung:2009bs,Anastasiou:2009kn} which allow one to derive validity
ranges on the kinematical variables.  Furthermore, in
Ref.\,\cite{Marzani:2008az}, the effects of the partonic high-energy
region on the total cross section have been studied by deriving the
leading behaviour in this limit.

A rather direct way to check the heavy-top limit is to evaluate formally
subleading terms. In this paper, we consider them for the purely virtual
corrections at \nnlo{}. While they do not correspond to a physical
quantity, they constitute an important gauge-invariant ingredient to the
full inclusive cross section. Note that at \nlo{}, the virtual
corrections are known in closed analytical form for arbitrary values of
$\mtop{}$~\cite{Harlander:2005rq,Anastasiou:2006hc,Aglietti:2006tp}.

Our approach is very similar to the calculation of the top mass
suppressed terms to the Higgs decay rate into gluons, described in
Ref.\,\cite{Schreck:2007um}. One might be tempted to use this result
obtained for the decay rate as an estimate of the effects for the gluon
fusion process. However, one should recall that the kinematics of the
two processes are very different. In particular, the top quark mass is
indeed the largest scale for the decay, so that the expansion in
$\mhiggs/\mtop$ remains within the radius of convergence. This is not 
the case for the higher order corrections to the gluon fusion production 
process involving real radiation of gluons and
quarks. The partonic center-of-mass energy $\sqrt{\hat s}$ can well
exceed the threshold value of $2\mtop$, and a series expansion in the
limit of large top mass becomes questionable. For the purely virtual
effects though, which are the subject of this paper, the partonic
center-of-mass energy is fixed to $\mhiggs$ which, according to the limits
derived from electro-weak precision fits, can safely be assumed to be
lighter than twice the top mass. They will therefore be a useful
ingredient for any possible treatment of the full hadronic cross
section, be it inclusive or exclusive.


\section{Method}

Sample diagrams that contribute to the virtual corrections to gluon
fusion at \lo{}, \nlo{}, and \nnlo{} are shown in \fig{fig::samples}.
An efficient and algorithmic procedure for evaluating them in terms of a
consistent expansion in $\mhiggs/\mtop$ is the well-known method of
asymptotic expansions~(see, e.g., Ref.\,\cite{Smirnov:2002pj}). In our
case, it expresses the original diagrams as a sum of convolutions of
massive vacuum with massless vertex integrals. The diagrammatic
representation of this procedure is shown for two particular diagrams in
Figs.\,\ref{fig::asympt1} and \ref{fig::asympt2}.


%
\newlength{\diasize}
\setlength{\diasize}{.22\textwidth}
\begin{figure}
  \begin{center}
    \begin{tabular}{ccc}
      \mbox{\includegraphics[width=\diasize]{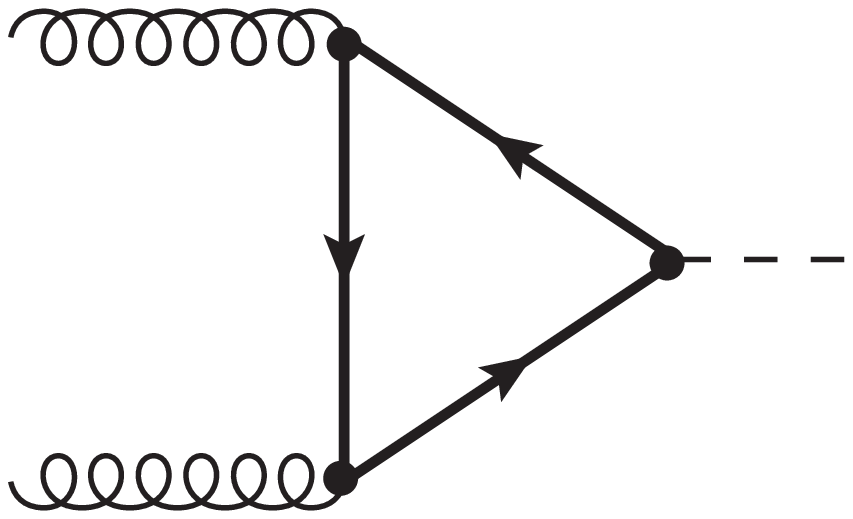}} 
      &\mbox{\includegraphics[width=\diasize]{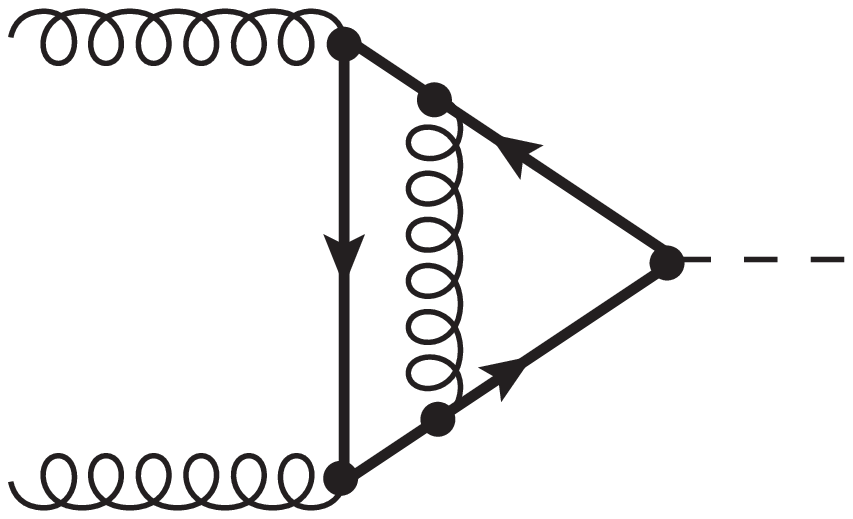}}
      &\mbox{\includegraphics[width=\diasize]{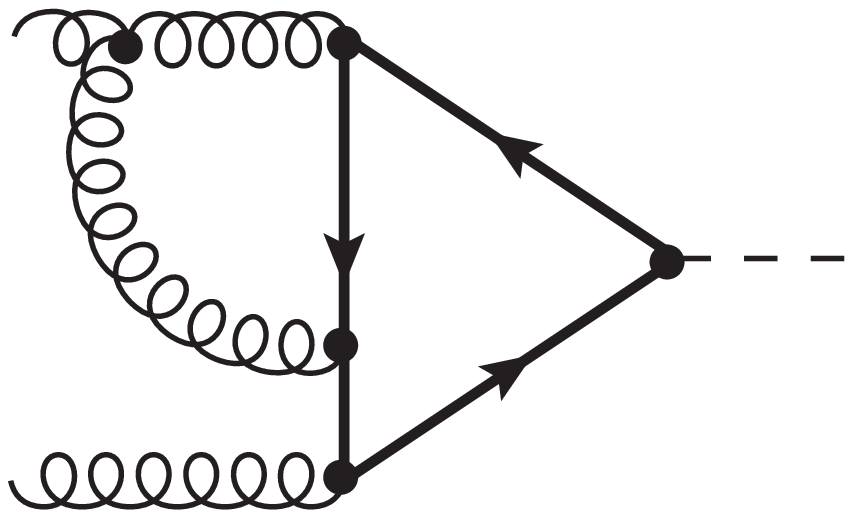}}\\
      (a) & (b) & (c)\\
      \mbox{\includegraphics[width=\diasize]{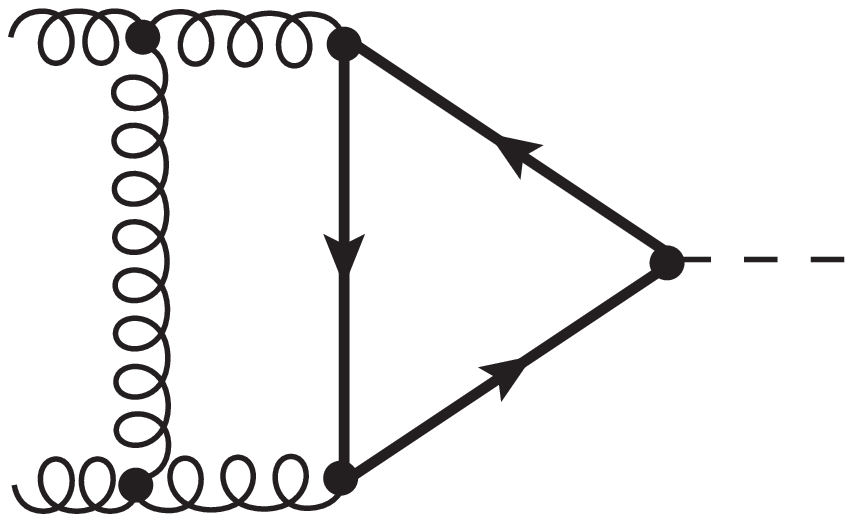}} 
      &\mbox{\includegraphics[width=\diasize]{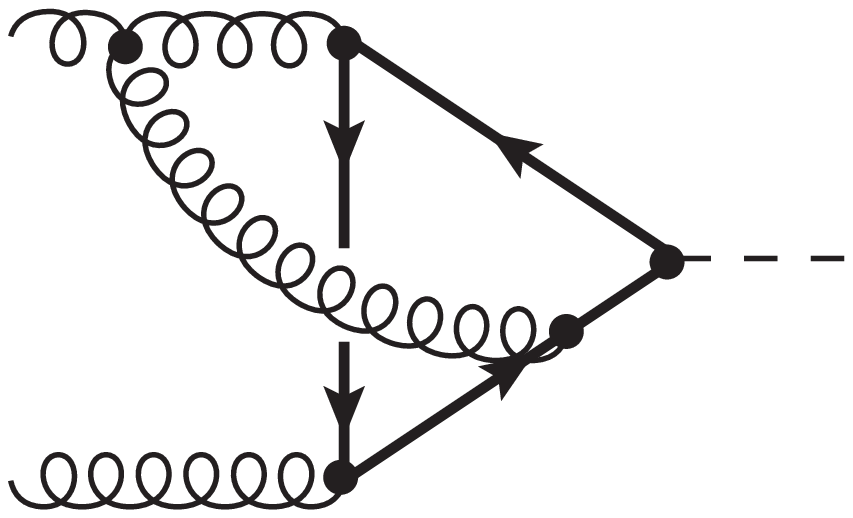}}
      &\mbox{\includegraphics[width=\diasize]{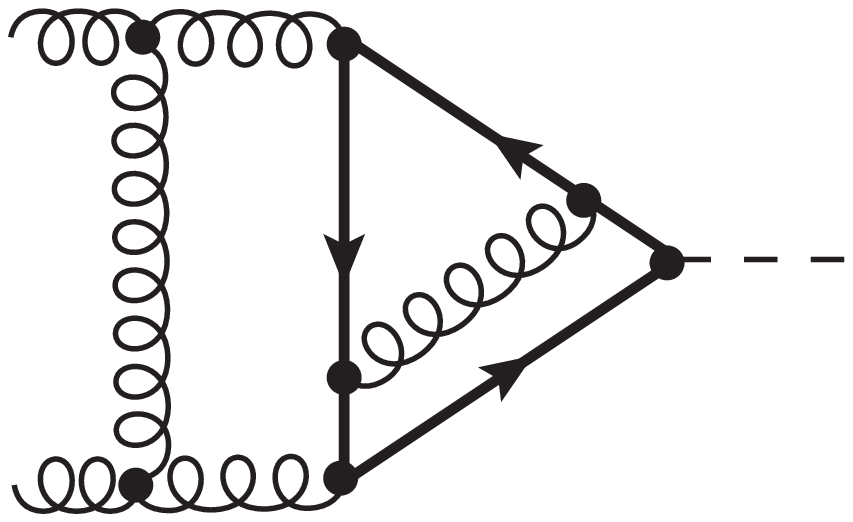}}\\
      (d) & (e) & (f)\\
      \mbox{\includegraphics[width=\diasize]{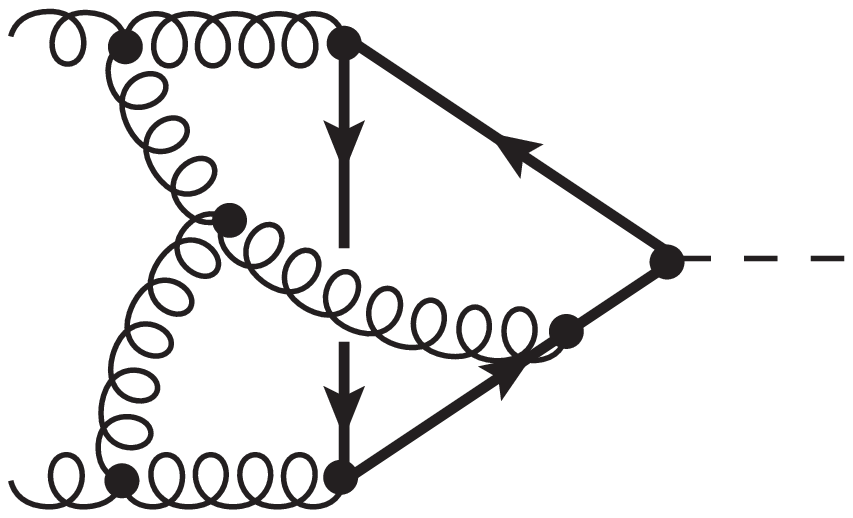}} 
      &\mbox{\includegraphics[width=\diasize]{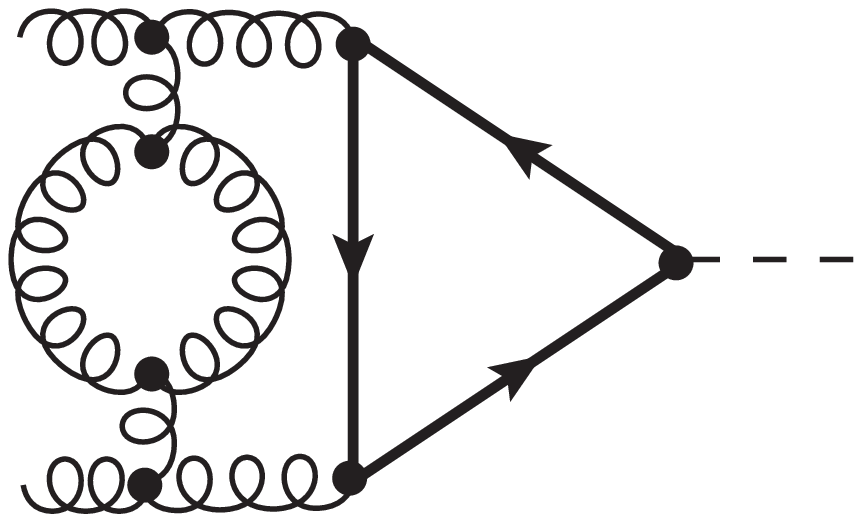}}
      &\mbox{\includegraphics[width=\diasize]{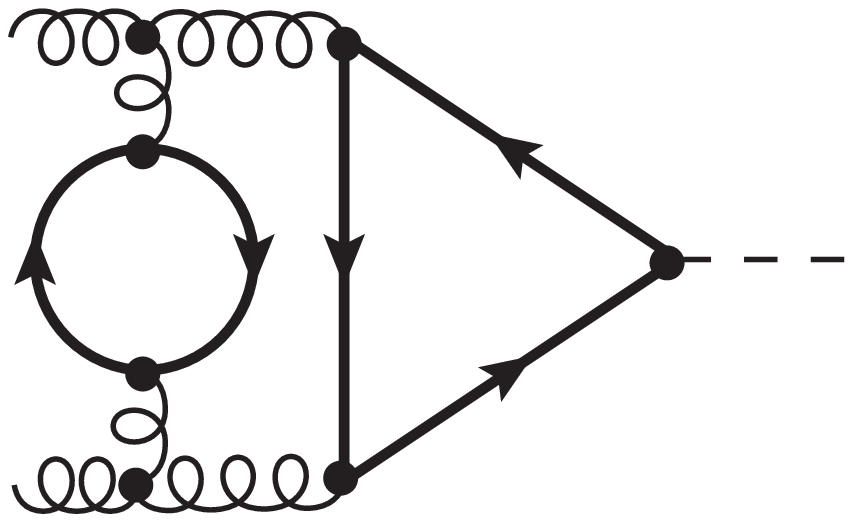}}\\
      (g) & (h) & (i)\\
    \end{tabular}
    \parbox{.9\textwidth}{
      \caption[]{\label{fig::samples}\sloppy Examples of Feynman
        diagrams contributing to the virtual corrections at \nnlo{} in
        the gluon fusion process. The solid lines denote top quarks, the
        springy lines are gluons, and the dashed line is the Higgs
        boson. In diagram (i), the bubble insertion can be a top quark
        or any other quark.  }}
  \end{center}
\end{figure}
%


%
\begin{figure}
  \begin{center}
    \begin{tabular}{c}
      \mbox{\includegraphics[width=\diasize]{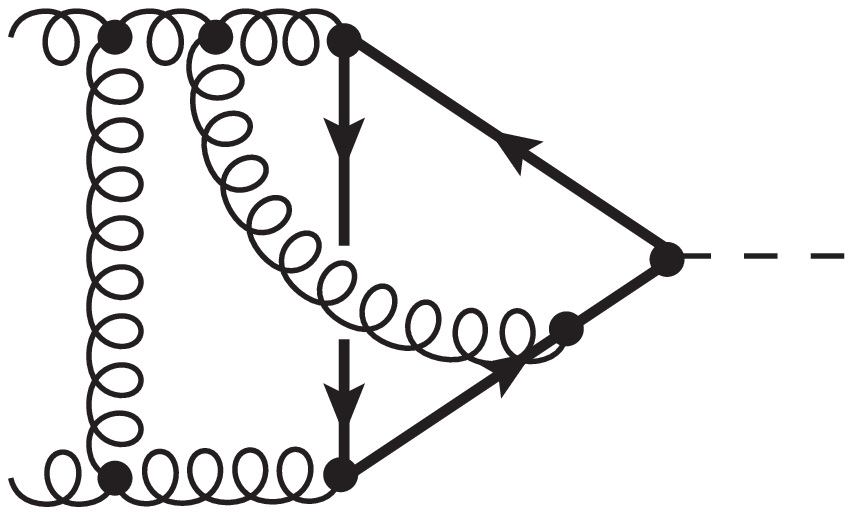}}\quad
      \raisebox{.27\diasize}{$\to$}\quad
      \mbox{\includegraphics[width=.97\diasize]{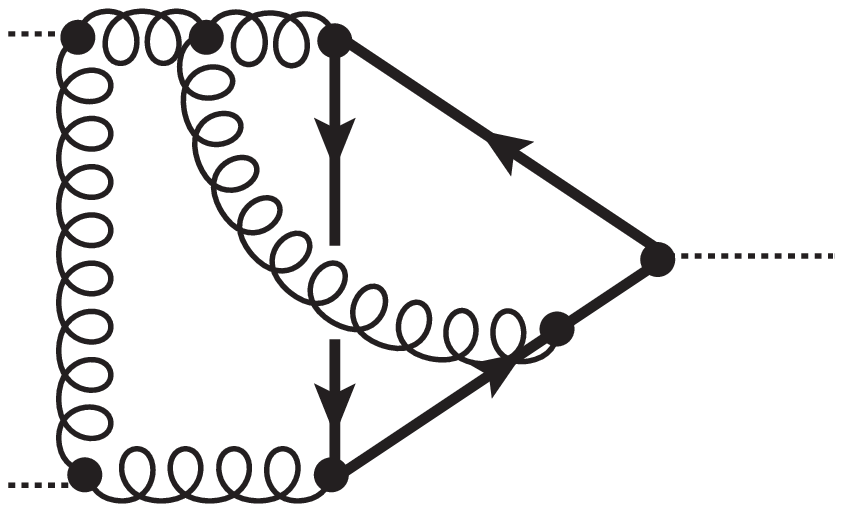}}\quad
      \raisebox{.27\diasize}{$\otimes$}
      \mbox{\includegraphics[width=.7\diasize]{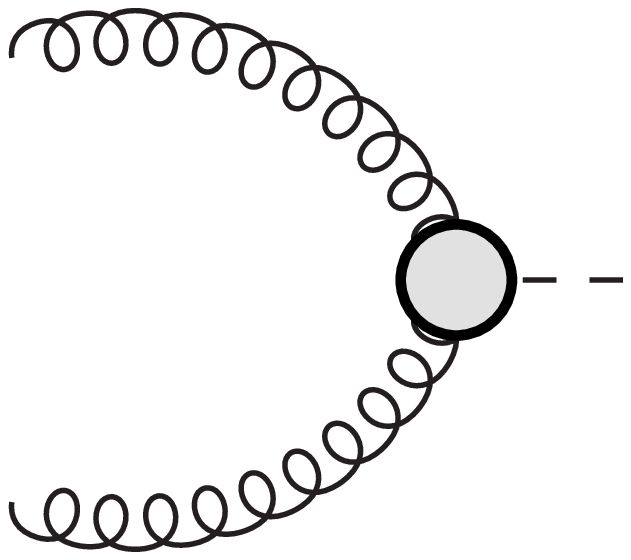}}\\[.5em]
      \raisebox{.27\diasize}{$+$}\quad
      \mbox{\includegraphics[width=.9\diasize]{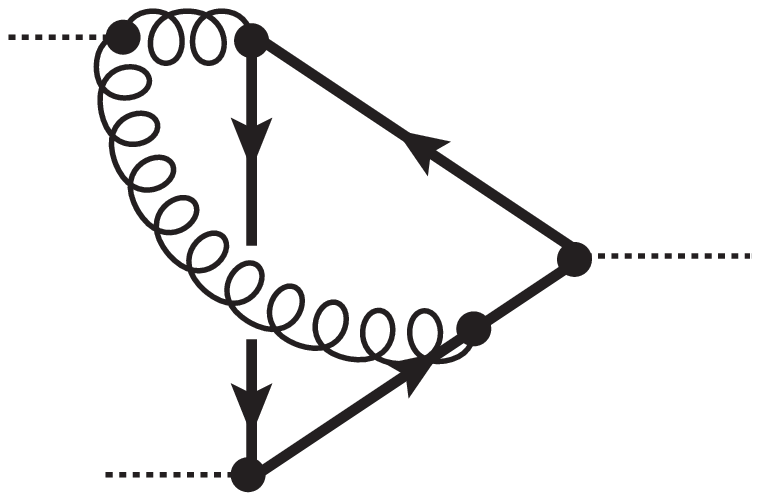}}\quad
      \raisebox{.27\diasize}{$\otimes$}
      \mbox{\includegraphics[width=.7\diasize]{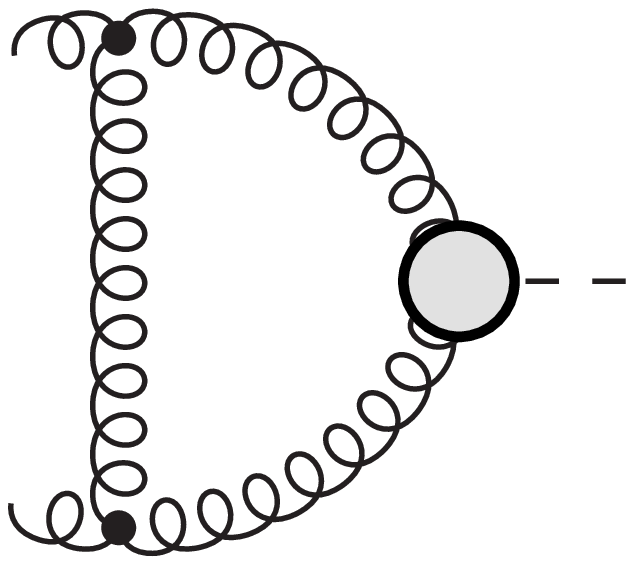}}
      \raisebox{.27\diasize}{$+$}\quad
      \mbox{\includegraphics[width=.7\diasize]{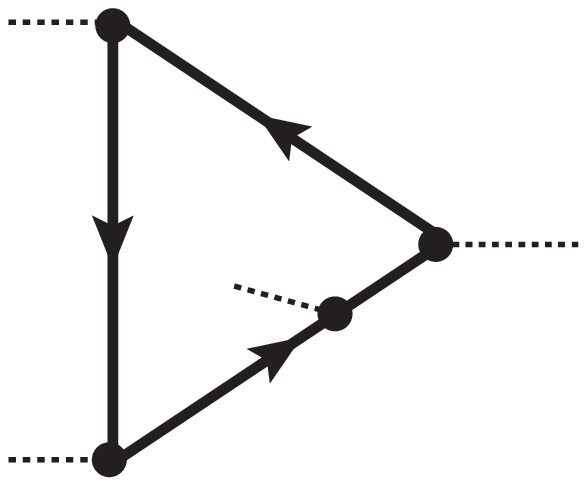}}\quad
      \raisebox{.27\diasize}{$\otimes$}
      \mbox{\includegraphics[width=.7\diasize]{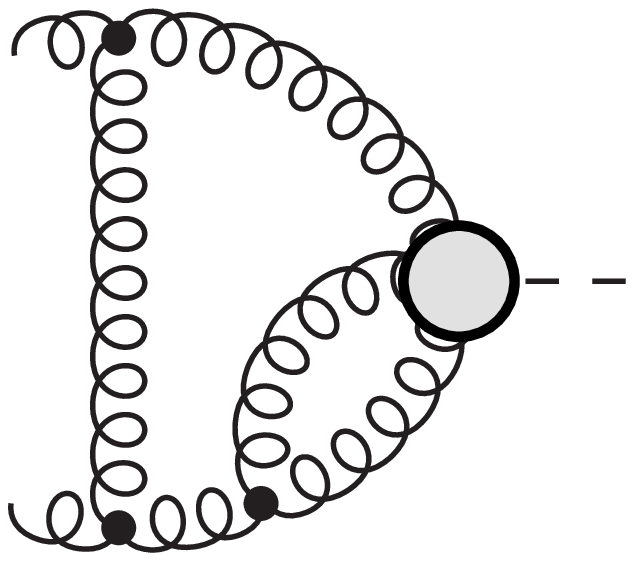}}
    \end{tabular}
    \parbox{.9\textwidth}{
      \caption[]{\label{fig::asympt1}\sloppy Diagrammatic representation
        for the asymptotic expansion of a particular Feynman diagram in
        the limit $\mhiggs^2\ll 4\mtop^2$. The diagrams left of
        $\otimes$ represent subdiagrams of the original diagram that are
        to be expanded in the momenta corresponding to the dotted
        external lines before the loop integration. In this way, it is
        apparent that the original integral, depending on $\mhiggs^2$
        and $\mtop^2$, is decomposed into products of ``tadpole''
        integrals with vanishing external momenta and massless vertex
        integrals. The shaded blob in the diagrams right of $\otimes$
        represents an effective vertex given by the result of the
        diagram left of $\otimes$ (for details of asymptotic expansions,
        see Ref.\cite{Smirnov:2002pj}, for example). The three terms
        right of ``$\to$'' are proportional to ${\cal N}_t^3$, ${\cal
          N}^2_t{\cal N}_h$, and ${\cal N}_t{\cal N}_h^2$, respectively
        (cf.~\eqn{eq::nn} below).  Subdiagrams without external mass
        scales are not shown.}}
  \end{center}
\end{figure}
%



%
\begin{figure}
  \begin{center}
    \begin{tabular}{c}
      \mbox{\includegraphics[width=\diasize]{figs/nnlo3.eps}}\quad
      \raisebox{.27\diasize}{$\to$}\\[.5em]
      \mbox{\includegraphics[width=\diasize]{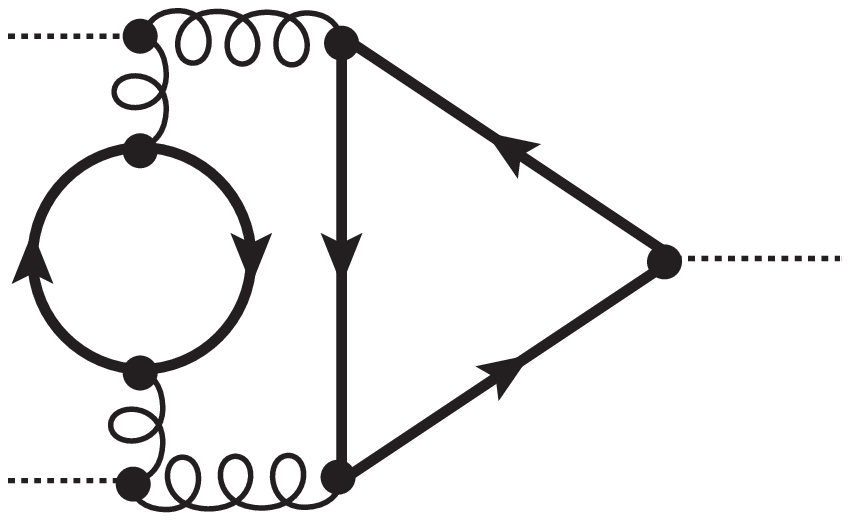}}\quad
      \raisebox{.27\diasize}{$\otimes$}
      \mbox{\includegraphics[width=.7\diasize]{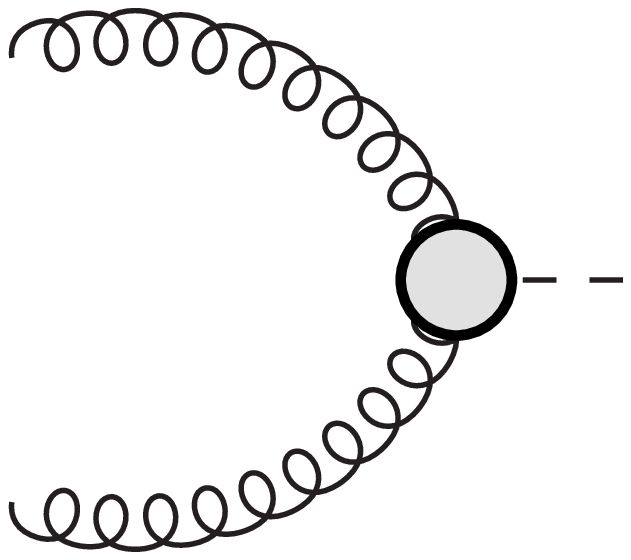}}\quad
      \raisebox{.27\diasize}{$+$}\quad
      \mbox{\includegraphics[width=1\diasize]{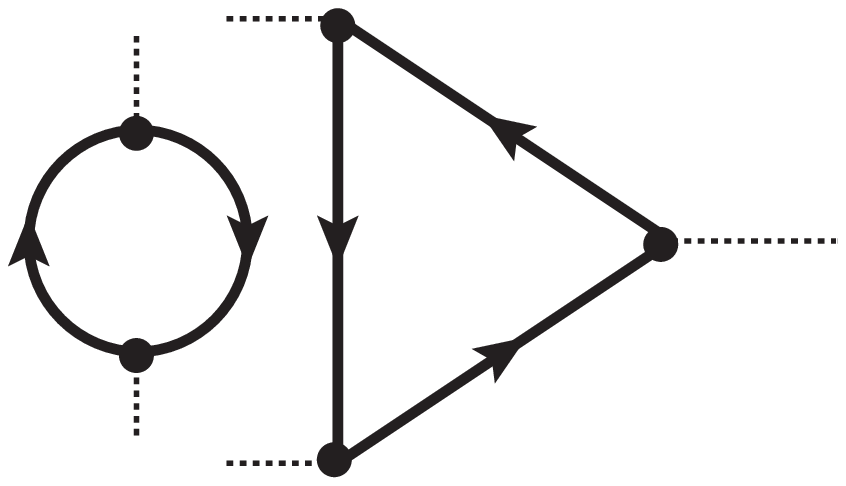}}\quad
      \raisebox{.27\diasize}{$\otimes$}
      \mbox{\includegraphics[width=.7\diasize]{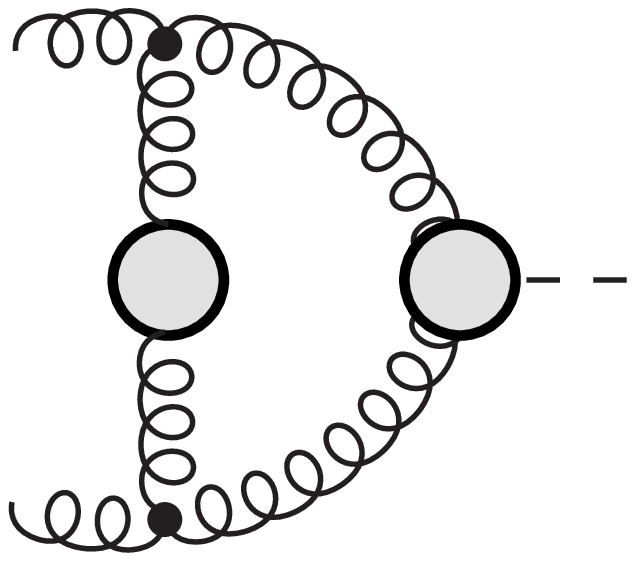}}
    \end{tabular}
    \parbox{.9\textwidth}{
      \caption[]{\label{fig::asympt2}\sloppy Diagrammatic representation
        for the asymptotic expansion of the Feynman diagram in
        \fig{fig::samples}~(i) in the limit $\mhiggs^2\ll 4\mtop^2$,
        when the bubble insertion is a top quark.  The second term on
        the r.h.s.\ is a source of the term $\sim\zeta_g^{(1),\bare}$ in
        \eqn{eq::h01} below. }}
  \end{center}
\end{figure}
%


We generate the diagrams with the help of {\tt qgraf}~\cite{Nogueira:1993ex} 
and pass them to {\tt q2e/exp}~\cite{Q2E,Harlander:1997zb}, which automatically
carries out the expansion. The resulting 1-, 2-, and 3-loop vacuum integrals
are evaluated by {\tt MATAD}~\cite{Steinhauser:2000ry}.  For the 1- and
2-loop vertex integrals we use the method of Ref.\,\cite{Baikov:2000jg}
by applying the relevant modifications~\cite{Harlander:2000mg} to {\tt
  MINCER}~\cite{Larin:1991fz}.

The colour and Lorentz structure of the physical amplitude is given by
\begin{equation}
\begin{split}
P^{ab}_{\mu\nu}(q_1,q_2) \equiv \delta^{ab}\left(
\dotprod{q_1}{q_2}\,g_{\mu\nu} - q_{1,\nu}q_{2,\mu}\right)\,,
\label{eq::proj}
\end{split}
\end{equation}
where $q_1^\mu$ and $q_2^\nu$ are the external gluon momenta, and $a$
and $b$ are the corresponding colour indices. We contract the amplitude with
$P^{ab}_{\mu\nu}(q_1,q_2)$ in order to arrive at a scalar expression in Lorentz and colour space.

Before the massive two- and three-loop integrals are passed to {\tt
  MATAD}, we need to eliminate any external momenta in their numerators
by appropriate decompositions into invariants, e.g.
\begin{equation}
\begin{split}
\int\dd^Dl\cdots\frac{(\dotprod{p_1}{l})(\dotprod{p_2}{l})}
       {\cdots(l^2-m^2)\cdots} =
       \frac{1}{D}(\dotprod{p_1}{p_2})\int\dd^Dl\cdots\frac{l^2}
            {\cdots(l^2-m^2)\cdots}\,,
\end{split}
\end{equation}
where the dots represent factors that are independent of $l$.

There are two diagrams at one-loop level, 23 at two-loop level, and 657
at three-loop level, and the calculation of the $1/\mtop^2$-suppressed
terms takes about $5\cdot 10^4$s, with the computationally most
expensive one shown in \fig{fig::asympt2}.


\section{Results}

Before we present the results, let us introduce some useful notation.
The renormalization scale $\mu$ appears in our calculation only through
the factors
\begin{equation}
\begin{split}
{\cal N}_h = e^{i\pi\ep}\left(\frac{\mu}{\mhiggs}\right)^{2\ep}{\cal N}\,,\qquad
{\cal N}_t = \left(\frac{\mu}{\mtop^{\bare}}\right)^{2\ep}{\cal N}\,,\qquad
{\cal N} = \exp[\ep(-\gamma_{\rm E} + \ln4\pi)]\,,
\label{eq::nn}
\end{split}
\end{equation}
with Euler's constant $\gamma_{\rm E} \approx 0.577216$\,. These
expressions are understood as their Laurent series in $\ep=(4-D)/2$,
where $D$ is the number of space-time dimensions used in the
calculation.

The perturbative coefficients typically contain the transcendental
numbers $\zeta_n\equiv \zeta(n)$, where $\zeta$ is Riemann's zeta
function. The particular values occurring here are
\begin{equation}
\begin{split}
\zeta_2 = \frac{\pi^2}{6} \approx  1.64493\,,\qquad
\zeta_3 \approx  1.20206\,,\qquad \zeta_4 = \frac{\pi^4}{90}\approx
1.08232\,.
\end{split}
\end{equation}

Throughout this paper, bare quantities are labeled by a superscript
``$\bare$''. Note that since the diagrams are evaluated with a spectrum
of six quark flavours, renormalization has to be performed accordingly.
To perform on-shell mass renormalization or conversion of $\alpha_s$ from the six-
to the five-flavour scheme one must keep the proper number of higher
order terms in $\ep$ due to the presence of infra-red
poles. Furthermore, in order to arrive at a
physical result, the external gluons must be renormalized on-shell.

For convenience, the number of light flavours $n_l$ is kept as a free
parameter; the physical case corresponds to $n_l=5$. The virtual cross section 
for the process $gg\to H$ can be written as
\begin{equation}
\begin{split}
\sigma_{\rm virt} &=
\frac{\pi}{576v^2}\frac{1}{(1-\ep)}\left(\apib\right)^2\delta(1-z)|{\cal
  N}_tH(\ep)\,h(\alpha_s^\bare)|^2\,,
\end{split}
\end{equation}
where
\begin{equation}
\begin{split}
 H(\ep) &= \Gamma(1+\ep)\bigg[ 1 +
 \frac{7}{120}\left(\frac{\mhiggs}{\mtop^\bare}\right)^2\left( 1 + \ep\right)
\\&\qquad
 + \frac{1}{168}\left(\frac{\mhiggs}{\mtop^\bare}\right)^4\left( 1 +
 \frac{3}{2}\ep +\frac{1}{2} \ep^2 \right) \bigg] +
 \order{\frac{\mhiggs^6}{\mtop^6}}\,.
\end{split}
\end{equation}
The amplitude is expanded in terms of a perturbative series:
\begin{equation}
\begin{split}
h(\alpha_s^\bare) &= 1 + \apib\,h^{(1)} + \left(\apib\right)^2\,h^{(2)} +
  \ldots\,,
\label{eq::amp}
\end{split}
\end{equation}
where the coefficients $h^{(n)}$ are functions of $\mhiggs$,
$\mtop^\bare$, and the renormalization scale $\mu$.  In our approach,
they take the form
\begin{equation}
\begin{split}
h^{(n)} &= h^{(n)}_0 + \left(\frac{\mhiggs}{\mtop^\bare}\right)^2 
h^{(n)}_2 + \left(\frac{\mhiggs}{\mtop^\bare}\right)^4 
h^{(n)}_4 + \ldots\,.
\label{eq::ampexp}
\end{split}
\end{equation}

The leading terms have been calculated in the framework of an effective
Lagrangian. However, for consistency, we present them here in a form
that is directly compatible with the mass suppressed terms to be
presented below:
\begin{equation}
\begin{split}
h^{(1)}_0 &= a^{(1)} + c^{(1)}\,,\\
h^{(2)}_0 &= a^{(2)} + c^{(2)} + a^{(1)}\,c^{(1)}
- \zeta_g^{(1),\bare} a^{(1)}\,,
\label{eq::h01}
\end{split}
\end{equation}
where
\begin{equation}
  \begin{split}
    a^{(1)} &= {\cal N}_h
    \bigg\{
    -\frac{3}{2\ep^2}
    + \frac{3}{4}\,\zeta_2 
    + \ep\,\bigg(
    - \frac{3}{2} 
    + \frac{7}{2}\,\zeta_3
    \bigg)
    + \ep^2\,\bigg(
    - \frac{9}{2}
    + \frac{141}{32}\,\zeta_4
    \bigg)
    \bigg\} + \order{\ep^3}\,,\\
    a^{(2)} &= {\cal N}_h^2 \bigg\{\frac{9}{8\ep^4} +
    \frac{1}{\ep^3}\,\bigg[ -\frac{33}{32} + \frac{1}{16}\,n_l \bigg] +
    \frac{1}{\ep^2}\,\bigg[ -\frac{67}{32} - \frac{9}{16}\,\zeta_2 +
      \frac{5}{48}\,n_l \bigg] \\&\quad + \frac{1}{\ep}\,\bigg[ \frac{17}{12}
      + \frac{99}{32}\,\zeta_2 - \frac{75}{16}\,\zeta_3 + n_l\,\bigg(
      -\frac{19}{72} - \frac{3}{16}\,\zeta_2 \bigg) \bigg] \\&\quad +
    \frac{5861}{288} + \frac{201}{32}\,\zeta_2 + \frac{11}{16}\,\zeta_3
    - \frac{189}{32}\,\zeta_4 + n_l\,\bigg[ -\frac{605}{216} -
      \frac{5}{16}\,\zeta_2 - \frac{7}{8}\,\zeta_3 \bigg] \bigg\}
    +\order{\ep}\,,
  \end{split}
  \label{eq::a2}
\end{equation}
are the perturbative coefficients of the effective Higgs-gluon vertex as
presented in Ref.\,\cite{Harlander:2000mg} which we quote here for the sake of
completeness. Furthermore, we find
\begin{equation}
\begin{split}
c_1^{(1)} &= 
{\cal N}_t\left[ \frac{3}{4} - \frac{11}{6}\ep + \left(\frac{17}{4} +
  \frac{3}{8}\zeta_2\right)\ep^2\right] + \order{\ep^3} \,,\\
c_2^{(2)} &= 
 {\cal N}_t^2\left[
  \frac{3}{32\ep^2} + \frac{1}{\ep}\left(\frac{241}{192} +
  \frac{5}{48}n_l\right) - \frac{4529}{1152} + \frac{3}{32}\zeta_2
  -\frac{73}{288}n_l\right] + \order{\ep}\,.
\end{split}
\end{equation}
It may be worth noting that $c^{(1)}$ and $c^{(2)}$ correspond to the
one- and two-loop results for the bare coefficient function of the
effective Lagrangian:\footnote{%
Renormalizing $C_1^\bare$ according to Ref.\,\cite{Steinhauser:2002rq}
(where $C_1^\bare$ is called $C_1^0$), one may derive the coefficient
function $C_1$ quoted in Eq.\,(3) of Ref.\,\cite{Harlander:2000mg}.}
\begin{equation}
\begin{split}
C_1^\bare &= -\frac{1}{3}\apib \Gamma(1+\ep)
\left[1 + \apib c_1^{(1)} + \left(\apib\right)^2 c_2^{(2)}
+ \order{\alpha_s^3}\right]\,.\\
\end{split}
\end{equation}

Finally,
\begin{equation}
\begin{split}
\zeta_g^{(1),\bare} = \frac{{\cal N}_t}{6}\left(
         \frac{1}{\ep}
          + \frac{\ep}{2}\zeta_2
          - \frac{\ep^2}{3}\zeta_3
          \right) + \order{\ep^3}
\end{split}
\end{equation}
is the 1-loop term of the bare decoupling constant for $\alpha_s$ for
the transition from $n_f=6$ to $n_l=5$ flavour \qcd{} (see, e.g.,
Ref.\,\cite{Chetyrkin:1997un}). The origin of the term involving
$\zeta_g^{(1),\bare}$ in \eqn{eq::h01} is the fact that the coefficients
$a^{(n)}$ in Ref.\,\cite{Harlander:2000mg} were evaluated in 5-flavour
\qcd{}, while the $h^{(n)}$ of \eqn{eq::h01} are based on 6-flavour
\qcd{}. Therefore, diagrams like the right-most one in
\fig{fig::asympt2} do not have a correspondence in the effective theory
calculation of Ref.\,\cite{Harlander:2000mg}.

The expressions presented so far correspond to known results and have
been included in this paper only for the sake of the reader's
convenience. They should facilitate any implementation of the newly
calculated terms to be presented below.  Besides that, they serve as a
useful check of our setup. It should be noted that in our approach, we
directly calculate the coefficients $h^{(n)}_m$, and the decomposition
into $a^{(n)}$ and $c^{(n)}$ is just for comparison to the literature.

The new results of this paper are the contributions to the virtual
3-loop amplitude that are formally suppressed by powers of
$\mhiggs/\mtop$. In the notation of Eqs.\,(\ref{eq::amp}) and
(\ref{eq::ampexp}), the first two subleading orders read\footnote{For
  the sake of brevity we insert {\abbrev SU(3)} colour factors.  The
  result for general colour factors can be obtained upon request from
  the authors.}
\begin{equation}
\begin{split}
h^{(1)}_2 &=
{\cal N}_t\bigg[-\frac{7}{60}\,\frac{1}{\ep} - \frac{79}{540}
+ \ep \left(\frac{37}{2400} - \frac{7}{120}\zeta_2\right)\\&\qquad
+ \ep^2 \left(-\frac{362063}{1944000} - \frac{79}{1080}\zeta_2
+ \frac{7}{180}\zeta_3\right)\bigg] + \order{\ep^3}\,,\\
h^{(2)}_2 &=
{\cal N}_t{\cal N}_h\bigg[ \frac{7}{40}\frac{1}{\ep^3}
+\frac{79}{360}\frac{1}{\ep^2}
-\frac{47}{1600}\frac{1}{\ep}
+\frac{588863}{1296000}
-\frac{7}{15}\zeta_3
\bigg]\\&\quad
+ {\cal N}^2_t\bigg[
\frac{1}{\ep^2}\left(-\frac{49}{1440}
+\frac{7}{720}\, n_l\right)
+ \frac{1}{\ep}\left(-\frac{703}{6480}
+\frac{919}{103680}\,n_l\right)
\\&\qquad
-\frac{255224167}{12441600}
-\frac{49}{1440}\zeta_2
+\frac{9556657}{552960}\zeta_3
+n_l\,\left(\frac{49729}{518400}
+\frac{7}{720}\zeta_2\right)
\bigg]  \\&\quad
+ {\cal N}_h^2\bigg[
\frac{1}{\ep}\left(-\frac{9}{320}
+ \frac{1441}{103680}\,n_l\right)
- \frac{559}{2560}
+ \frac{36377}{311040}n_l
\bigg] + \order{\ep}\,,
\end{split}
\end{equation}
\begin{equation}
\begin{split}
h^{(1)}_4 &= {\cal N}_t\bigg[-\frac{857}{50400}\,\frac{1}{\ep} -
  \frac{3301}{113400} + \ep \left(-\frac{1064509}{76204800} -
  \frac{857}{100800}\zeta_2\right)\\&\qquad + \ep^2
  \left(-\frac{506339}{21952000} - \frac{3301}{226800}\zeta_2 +
  \frac{857}{151200}\zeta_3\right)\bigg] + \order{\ep^3} \,,\\
h^{(2)}_4 &=
{\cal N}_t{\cal N}_h\bigg[ \frac{857}{33600}\frac{1}{\ep^3}
+\frac{3301}{75600}\frac{1}{\ep^2}
+\frac{1175389}{50803200}\frac{1}{\ep}
+\frac{240009257}{3556224000}
-\frac{857}{12600}\zeta_3
\bigg]\\&\quad
+ {\cal N}^2_t\bigg[
\frac{1}{\ep^2}\left(\frac{22801}{1209600}
+\frac{857}{604800}\, n_l\right)
+ \frac{1}{\ep}\left(\frac{28471}{1088640}
+\frac{94907}{29030400}\,n_l\right)
\\&\qquad
-\frac{5277060458353}{1170505728000}
+\frac{22801}{1209600}\zeta_2
+\frac{9358312739}{2477260800}\zeta_3
\\&\qquad\quad
+n_l\,\left(\frac{13897721}{1143072000}
+\frac{857}{604800}\zeta_2\right)
\bigg]  \\&\quad
+ {\cal N}_h^2\bigg[
\frac{1}{\ep}\left(\frac{87}{89600}
+ \frac{80231}{87091200}\,n_l\right)
+ \frac{1481}{153600}
+ \frac{781}{102060}n_l
\bigg] + \order{\ep}\,,\\
\end{split}
\end{equation}
where $h_2^{(1)}$ is the two-loop result for which the $\ep^{-1}$ and
$\ep^0$ terms can be compared to Ref.\,\cite{Dawson:1993qf}, with full
agreement, of course.  Note that since the leading order
amplitude has been factored out in terms of $H(\ep)$, the \nlo{} and
\nnlo{} expressions for $h$ start only at $\order{1/\ep}$ and
$\order{1/\ep^3}$ respectively. The leading poles are thus fully
determined by the leading terms in $1/\mtop$.

There are a number of checks that we can perform on our result: (i)
calculating the amplitude in an arbitrary covariant gauge, we find it to
be independent of the gauge parameter; (ii)~replacing the projector of
\eqn{eq::proj} by $\delta^{ab}q_{1,\mu}q_{2,\nu}$ leads to a vanishing
result, which also checks gauge invariance; (iii) the poles of order
$\alpha_s^{n+1}/\ep^k$, $k=1,\ldots,2n$ were checked against the general
formula of Ref.\,\cite{Catani:1998bh} (with additional input from
Ref.\,\cite{Glover:2003cm} for the $\alpha_s^3/\ep$ terms; see also
Ref.\,\cite{Becher:2009cu:2009qa}) and found to be in full agreement.

Another observation is that in the ratio of the amplitudes taken at
time-like and space-like momenta, all mass effects cancel and the result
is the one given by the general formula of
Ref.\,\cite{Magnea:1990zb,Moch:2005tm}. Specifically,
\begin{equation}
\begin{split}
\left|\frac{\hat h(M_H^2)}{\hat h(-M_H^2)}\right|^2
=
\left|\frac{a(M_H^2)}{a(-M_H^2)}\right|^2\,,
\label{eq::timespaceratio}
\end{split}
\end{equation}
where
\begin{equation}
\begin{split}
\hat h(q^2) &= Z_3^{\rm OS} H(\ep)
h(\alpha_s^\bare)\left|_{\mhiggs^2=q^2}\right.\,,
\end{split}
\end{equation}
with the on-shell gluon renormalization constant $Z_3^{\rm
  OS}$~\cite{Chetyrkin:1997un} (see also~Ref.~\cite{Czakon:2007ej}). The
ratio on the left-hand side of \eqn{eq::timespaceratio} can also be
found in Ref.\,\cite{Harlander:2000mg}.  It is understood that the bare
quantities $\mtop^\bare$ and $\alpha_s^\bare$ in
\eqn{eq::timespaceratio} are expressed in terms of their renormalized
values~\cite{Melnikov:2000zc}.

\section{Conclusions}

The top mass suppressed terms for the virtual corrections to Standard
Model Higgs production in gluon fusion were presented through three-loop
order. They are an ingredient for the inclusive \nnlo{} \qcd{} cross
section. We have subjected the result to various checks and found full
confirmation.

The next steps towards the top mass effects in the Higgs production
cross section will be the evaluation of the mass suppressed terms in the
real radiation amplitudes.  We defer this problem to a forthcoming
publication.

Upon completion of this paper, we became aware of a similar
calculation~\cite{Pak:2009}. We have compared our results and found full
agreement.

\paragraph{Acknowledgments.}
This work has been supported by {\it Deutsche Forschungsgemeinschaft},
contract HA~2990/3-1 and the Helmholtz Alliance {\it Physics at the
  Terascale}. We would like to thank A.~Pak, M.~Rogal, and
M.~Steinhauser for useful comments, as well as M.~Czakon for
enlightening discussions concerning the renormalization of the
amplitude.




\def\app#1#2#3{{\it Act.~Phys.~Pol.~}\jref{\bf B #1}{#2}{#3}}
\def\apa#1#2#3{{\it Act.~Phys.~Austr.~}\jref{\bf#1}{#2}{#3}}
\def\annphys#1#2#3{{\it Ann.~Phys.~}\jref{\bf #1}{#2}{#3}}
\def\cmp#1#2#3{{\it Comm.~Math.~Phys.~}\jref{\bf #1}{#2}{#3}}
\def\cpc#1#2#3{{\it Comp.~Phys.~Commun.~}\jref{\bf #1}{#2}{#3}}
\def\epjc#1#2#3{{\it Eur.\ Phys.\ J.\ }\jref{\bf C #1}{#2}{#3}}
\def\fortp#1#2#3{{\it Fortschr.~Phys.~}\jref{\bf#1}{#2}{#3}}
\def\ijmpc#1#2#3{{\it Int.~J.~Mod.~Phys.~}\jref{\bf C #1}{#2}{#3}}
\def\ijmpa#1#2#3{{\it Int.~J.~Mod.~Phys.~}\jref{\bf A #1}{#2}{#3}}
\def\jcp#1#2#3{{\it J.~Comp.~Phys.~}\jref{\bf #1}{#2}{#3}}
\def\jetp#1#2#3{{\it JETP~Lett.~}\jref{\bf #1}{#2}{#3}}
\def\jphysg#1#2#3{{\small\it J.~Phys.~G~}\jref{\bf #1}{#2}{#3}}
\def\jhep#1#2#3{{\small\it JHEP~}\jref{\bf #1}{#2}{#3}}
\def\mpl#1#2#3{{\it Mod.~Phys.~Lett.~}\jref{\bf A #1}{#2}{#3}}
\def\nima#1#2#3{{\it Nucl.~Inst.~Meth.~}\jref{\bf A #1}{#2}{#3}}
\def\npb#1#2#3{{\it Nucl.~Phys.~}\jref{\bf B #1}{#2}{#3}}
\def\nca#1#2#3{{\it Nuovo~Cim.~}\jref{\bf #1A}{#2}{#3}}
\def\plb#1#2#3{{\it Phys.~Lett.~}\jref{\bf B #1}{#2}{#3}}
\def\prc#1#2#3{{\it Phys.~Reports }\jref{\bf #1}{#2}{#3}}
\def\prd#1#2#3{{\it Phys.~Rev.~}\jref{\bf D #1}{#2}{#3}}
\def\pR#1#2#3{{\it Phys.~Rev.~}\jref{\bf #1}{#2}{#3}}
\def\prl#1#2#3{{\it Phys.~Rev.~Lett.~}\jref{\bf #1}{#2}{#3}}
\def\pr#1#2#3{{\it Phys.~Reports }\jref{\bf #1}{#2}{#3}}
\def\ptp#1#2#3{{\it Prog.~Theor.~Phys.~}\jref{\bf #1}{#2}{#3}}
\def\ppnp#1#2#3{{\it Prog.~Part.~Nucl.~Phys.~}\jref{\bf #1}{#2}{#3}}
\def\rmp#1#2#3{{\it Rev.~Mod.~Phys.~}\jref{\bf #1}{#2}{#3}}
\def\sovnp#1#2#3{{\it Sov.~J.~Nucl.~Phys.~}\jref{\bf #1}{#2}{#3}}
\def\sovus#1#2#3{{\it Sov.~Phys.~Usp.~}\jref{\bf #1}{#2}{#3}}
\def\tmf#1#2#3{{\it Teor.~Mat.~Fiz.~}\jref{\bf #1}{#2}{#3}}
\def\tmp#1#2#3{{\it Theor.~Math.~Phys.~}\jref{\bf #1}{#2}{#3}}
\def\yadfiz#1#2#3{{\it Yad.~Fiz.~}\jref{\bf #1}{#2}{#3}}
\def\zpc#1#2#3{{\it Z.~Phys.~}\jref{\bf C #1}{#2}{#3}}
\def\ibid#1#2#3{{ibid.~}\jref{\bf #1}{#2}{#3}}
\def\otherjournal#1#2#3#4{{\it #1}\jref{\bf #2}{#3}{#4}}

\newcommand{\jref}[3]{{\bf #1} (#2) #3}
\newcommand{\bibentry}[4]{#1, #3\ifthenelse{\equal{#4}{}}{}{, }#4.}
\newcommand{\hepph}[1]{{\tt [hep-ph/#1]}}
\newcommand{\arxiv}[2]{{\tt [arXiv:#1]}}



\begin{thebibliography}{99}
%
%

\bibitem{Dawson:1990zj}
\bibentry{S.~Dawson}
{Radiative corrections to Higgs boson production}
{\npb{359}{1991}{283}}
{}

\bibitem{Djouadi:1991tk}
\bibentry{A.~Djouadi, M.~Spira, P.M.~Zerwas}
{Production of Higgs bosons in proton colliders: {\abbrev QCD} corrections}
{\plb{264}{1991}{440}}
{}

\bibitem{Graudenz:1992pv}
\bibentry{D.~Graudenz, M.~Spira, P.M.~Zerwas}
{{\abbrev QCD} corrections to Higgs boson production at proton-proton
colliders}
{\prl{70}{1993}{1372}}
{}

\bibitem{Spira:1993bb}
\bibentry{M.~Spira, A.~Djouadi, D.~Graudenz, P.M.~Zerwas}
{SUSY Higgs production at proton colliders}
{\plb{318}{1993}{347}}
{}

\bibitem{Harlander:2002wh}
\bibentry{R.V.~Harlander and W.B.~Kilgore}
{Next-to-next-to-leading order Higgs production at hadron colliders}
{\prl{88}{2002}{201801}}
{\hepph{0201206}}

\bibitem{Anastasiou:2002yz}
\bibentry{C.~Anastasiou and K.~Melnikov}
{Higgs boson production at hadron colliders in {\abbrev NNLO QCD}}
{\npb{646}{2002}{220}}
{\hepph{0207004}}

\bibitem{Ravindran:2003um}
\bibentry{V.~Ravindran, J.~Smith, W.L.~van Neerven}
{{\abbrev NNLO} corrections to the total cross section for Higgs boson
  production  in hadron hadron collisions}
{\npb{665}{2003}{325}}
{\hepph{0302135}}

\bibitem{deFlorian:2009hc}
\bibentry{D.~de Florian and M.~Grazzini}
{Higgs production through gluon fusion: updated cross sections at the
 Tevatron and the LHC}
{\plb{674}{2009}{291}}
{\arxiv{0901.2427}{hep-ph}}

\bibitem{Anastasiou:2008tj}
\bibentry{C.~Anastasiou, R.~Boughezal, F.~Petriello}
{Mixed QCD-electroweak corrections to Higgs boson production in gluon
fusion}
{\jhep{0904}{2009}{003}}
{\arxiv{0811.3458}{hep-ph}}

\bibitem{Dawson:1993qf}
\bibentry{S.~Dawson and R.~Kauffman}
{QCD corrections to Higgs boson production: nonleading terms in the heavy
 quark limit}
{\prd{49}{1994}{2298}}
{\hepph{9310281}}

\bibitem{Kramer:1996iq}
\bibentry{M.~Kr\"amer, E.~Laenen, M.~Spira}
{Soft gluon radiation in Higgs boson production at the {\abbrev LHC}}
{\npb{511}{1998}{523}}
{\hepph{9611272}}

\bibitem{Harlander:2003xy}
\bibentry{R.~Harlander}
{Supersymmetric Higgs production at the Large Hadron Collider}
%
{\hepph{0311005}}
{}

\bibitem{Catani:2003zt}
\bibentry{S.~Catani, D.~de Florian, M.~Grazzini, P.~Nason}
{Soft-gluon resummation for Higgs boson production at hadron colliders}
{\jhep{0307}{2003}{028}}
{\hepph{0306211}}

\bibitem{Ahrens:2008qu:2008nc}
\bibentry{V.~Ahrens, T.~Becher, M.~Neubert, L.~L.~Yang}
{Origin of the Large Perturbative Corrections to Higgs Production at Hadron
 Colliders}
{\prd{79}{2009}{033013}, \arxiv{0808.3008}{hep-ph}; \arxiv{0809.4283}{hep-ph}}
{}

\bibitem{DelDuca:2001eu:2001fn:2003ba}
\bibentry{V.~Del Duca, W.~Kilgore, C.~Oleari, C.~Schmidt, D.~Zeppenfeld}
{H + 2 jets via gluon fusion}
{\prl{87}{2001}{122001}, \hepph{0105129}; \npb{616}{2001}{367}, 
\hepph{0108030}; \prd{67}{2003}{073003}; \hepph{0301013}}
{}

\bibitem{Keung:2009bs}
\bibentry{W.Y.~Keung and F.~Petriello}
{Electroweak and finite quark-mass effects on the Higgs boson transverse
momentum distribution}
{\arxiv{0905.2775}{hep-ph}}
{}

\bibitem{Anastasiou:2009kn}
\bibentry{C.~Anastasiou, S.~Bucherer, Z.~Kunszt}
{HPro: A NLO Monte-Carlo for Higgs production via gluon fusion with finite
heavy quark masses}
{\arxiv{0907.2362}{hep-ph}}
{}

\bibitem{Marzani:2008az}
\bibentry{S.~Marzani, R.D.~Ball, V.~Del Duca, S.~Forte, A.~Vicini}
{Higgs production via gluon-gluon fusion with finite top mass beyond
 next-to-leading order}
{\npb{800}{2008}{127}}
{\arxiv{0801.2544}{hep-ph}}

\bibitem{Harlander:2005rq}
\bibentry{R.~Harlander and P.~Kant}
{Higgs production and decay: Analytic results at next-to-leading order QCD}
{\jhep{0512}{2005}{015}}
{\hepph{0509189}}

\bibitem{Anastasiou:2006hc}
\bibentry{C.~Anastasiou, S.~Beerli, S.~Bucherer, A.~Daleo, Z.~Kunszt}
{Two-loop amplitudes and master integrals for the production of a Higgs
 boson via a massive quark and a scalar-quark loop}
{\jhep{0701}{2007}{082}}
{\hepph{0611236}}

\bibitem{Aglietti:2006tp}
\bibentry{U.~Aglietti, R.~Bonciani, G.~Degrassi, A.~Vicini}
{Analytic results for virtual QCD corrections to Higgs production and
 decay}
{\jhep{0701}{2007}{021}}
{\hepph{0611266}}

\bibitem{Schreck:2007um}
\bibentry{M.~Schreck and M.~Steinhauser}
{Higgs decay to gluons at NNLO}
{\plb{655}{2007}{148}}
{\arxiv{0708.0916}{hep-ph}}

\bibitem{Smirnov:2002pj}
\bibentry{V.A.~Smirnov}
{Applied asymptotic expansions in momenta and masses}
{\otherjournal{Springer Tracts Mod.\ Phys.}{177}{2002}{1}}
{}

\bibitem{Nogueira:1993ex}
\bibentry{P.~Nogueira}
{Automatic Feynman graph generation}
{\jcp{105}{1993}{279}}
{}

\bibitem{Q2E}
\bibentry{T.~Seidensticker}
{Using {\tt q2e} and {\tt exp}}
{Universit\"at Karlsruhe, 2002 (unpublished)}
{}

\bibitem{Harlander:1997zb}
\bibentry{R.~Harlander, T.~Seidensticker, M.~Steinhauser}
{Corrections of ${\cal O}(\alpha \alpha_s)$ to the decay of the $Z$ boson  
into bottom quarks}
{\plb{426}{1998}{125}}
{\hepph{9712228}}

\bibitem{Steinhauser:2000ry}
\bibentry{M.~Steinhauser}
{{\tt MATAD}: A program package for the computation of massive tadpoles}
{\cpc{134}{2001}{335}}
{\hepph{0009029}}

\bibitem{Baikov:2000jg}
\bibentry{P.A.~Baikov and V.A.~Smirnov}
{Equivalence of recurrence relations for Feynman 
integrals with the same total number of external and loop momenta}
{\plb{477}{2000}{367}}
{\hepph{0001192}}

\bibitem{Harlander:2000mg}
\bibentry{R.V.~Harlander}
{Virtual corrections to $g g \to H$ to two loops in the heavy top limit}
{\plb{492}{2000}{74}}
{\hepph{0007289}}

\bibitem{Larin:1991fz}
\bibentry{S.A.~Larin, F.V.~Tkachov, J.A.~Vermaseren}
{The Form Version Of Mincer}
{NIKHEF-H-91-18}
{}

\bibitem{Steinhauser:2002rq}
\bibentry{M.~Steinhauser}
{Results and techniques of multi-loop calculations}
{\pr{364}{2002}{247}}
{\hepph{0201075}}

\bibitem{Chetyrkin:1997un}
\bibentry{K.G.~Chetyrkin, B.A.~Kniehl, M.~Steinhauser}
{Decoupling relations to $\order{\alpha_s^3}$ and their connection to
low-energy theorems}
{\npb{510}{1998}{61}}
{\hepph{9708255}}

\bibitem{Catani:1998bh}
\bibentry{S.~Catani}
{The singular behaviour of {\abbrev QCD} amplitudes at two-loop order}
{\plb{427}{1998}{161}}
{\hepph{9802439}}

\bibitem{Glover:2003cm}
\bibentry{E.W.N.~Glover and M.E.~Tejeda-Yeomans}
{Two-loop QCD helicity amplitudes for massless quark-massless gauge boson
scattering}
{\jhep{0306}{2003}{033}}
{\hepph{0304169}}

\bibitem{Becher:2009cu:2009qa}
\bibentry{T.~Becher and M.~Neubert}
{Infrared singularities of scattering amplitudes in perturbative QCD}
{\prl{102}{2009}{162001}, \arxiv{0901.0722}{hep-ph}; \jhep{0906}{2009}{081},
\arxiv{0901.0722}{hep-ph}}
{}

\bibitem{Magnea:1990zb}
\bibentry{L.~Magnea and G.~Sterman}
{Analytic continuation of the Sudakov form-factor in QCD}
{\prd{42}{1990}{4222}}
{}

\bibitem{Moch:2005tm}
\bibentry{S.~Moch, J.A.M.~Vermaseren, A.~Vogt}
{Three-loop results for quark and gluon form factors}
{\plb{625}{2005}{245}}
{\hepph{0508055}}

\bibitem{Czakon:2007ej}
\bibentry{M.~Czakon, A.~Mitov, S.~Moch}
{Heavy-quark production in massless quark scattering at two loops in QCD}
{\plb{651}{2007}{147}}
{\arxiv{0705.1975}{hep-ph}}

\bibitem{Melnikov:2000zc}
\bibentry{K.~Melnikov and T.~van Ritbergen}
{The three-loop on-shell renormalization of QCD and QED}
{\npb{591}{2000}{515}}
{\hepph{0005131}}

\bibitem{Pak:2009}
\bibentry{A.~Pak, M.~Rogal, M.~Steinhauser}
{Virtual three-loop corrections to Higgs boson production in
gluon fusion for finite top quark mass}
{\arxiv{0907.2998}{hep-ph}}
{}

\end{thebibliography}
\end{document}